\documentclass[compsoc,conference,a4paper,10pt,times]{IEEEtran}
\IEEEoverridecommandlockouts
\usepackage{cite}
\usepackage{amsmath,amssymb,amsfonts}
\usepackage{algorithmic}
\usepackage{graphicx}
\usepackage{textcomp}
\usepackage{xcolor}
\usepackage[colorlinks=true,urlcolor=black]{hyperref}
\usepackage{listings}
\usepackage{subfiles}
\usepackage{float}
\usepackage{MnSymbol}%
\usepackage{wasysym}%
\usepackage{paralist}
\usepackage{booktabs}
\usepackage{caption}
\usepackage{verbatim}
\usepackage{xspace}
\usepackage{xurl}  
\usepackage{spverbatim}
\usepackage{enumitem}
\usepackage{diagbox} 
\usepackage{threeparttable} 

\hypersetup{
    colorlinks,
    linkcolor={red!50!black},
    citecolor={blue!50!black},
    urlcolor={blue!80!black}
}

\newcommand{\system}{ARVO\xspace}
\newcommand{\dataset}{ARVO dataset\xspace}

\usepackage{pifont}
\usepackage{tikz}     

\newcommand{\cmark}{\textcolor{green}{\ding{51}}}
\newcommand{\xmark}{\textcolor{red}{\ding{55}}}
\newcommand{\hmark}{
  \textcolor{green}{\ding{51}}%
  \kern-0.62em\raisebox{0.52ex}{\scriptsize\textcolor{red}{\ding{55}}}%
}
\def\BibTeX{{\rm B\kern-.05em{\sc i\kern-.025em b}\kern-.08em
    T\kern-.1667em\lower.7ex\hbox{E}\kern-.125emX}}
\begin{document}

\title{\Large \bf ARVO: Atlas of Reproducible Vulnerabilities for Open-Source Software}

\author{
\IEEEauthorblockN{
Xiang Mei\IEEEauthorrefmark{1},
Jordi Del Castillo\IEEEauthorrefmark{2},
Pulkit Singh Singaria\IEEEauthorrefmark{1},
Haoran Xi\IEEEauthorrefmark{2}\\
Abdelouahab Benchikh\IEEEauthorrefmark{1},
Tiffany Bao\IEEEauthorrefmark{1},
Ruoyu Wang\IEEEauthorrefmark{1},
Yan Shoshitaishvili\IEEEauthorrefmark{1}\\
Adam Doup\'e\IEEEauthorrefmark{1},
Hammond Pearce\IEEEauthorrefmark{3},
Brendan Dolan-Gavitt\IEEEauthorrefmark{4}
}
\IEEEauthorblockA{
\IEEEauthorrefmark{1}Arizona State University,
\IEEEauthorrefmark{2}New York University,
\IEEEauthorrefmark{3}University of New South Wales,
\IEEEauthorrefmark{4}XBOW\\
\IEEEauthorrefmark{1}\{n132, psingari, tbao, fishw, yans, doupe\}@asu.edu,
am.benchikh@esi-sba.dz\\
\IEEEauthorrefmark{2}\{jordi.d, hx759\}@nyu.edu,\quad
\IEEEauthorrefmark{3}hammond.pearce@unsw.edu.au,\quad
\IEEEauthorrefmark{4}moyix@xbow.com
}
}

\maketitle

\begin{abstract}

Achieving reproducibility, quantity, and diversity in vulnerability
datasets has long been viewed as an inherent three-way trade-off, where
improving one dimension often comes at the cost of the others. 
In practice, reproducibility has been the dimension most often neglected.
This has limited what can be automatically extracted from historical bug datasets, and has reduced their utility for downstream security research.

In this work, we propose a method to produce a new security dataset which ensures reproducibility for diverse vulnerabilities
at scale by identifying the key obstacles to large-scale bug reproduction
and addressing them with general solutions. Using this method, we introduce full reproducibility to the
largest open source software vulnerability dataset (OSS-Fuzz) and construct
the ARVO dataset (an \underline{A}tlas of \underline{R}eproducible
\underline{V}ulnerabilities in \underline{O}pen-source software). ARVO is a large-scale dataset consisting of over 6{,}100 real-world
vulnerabilities across 311 projects. Focusing on reproducibility, ARVO differs
from existing datasets by providing each vulnerability in a form that can
be consistently rebuilt, triggered, and analyzed across versions.
Reproducibility also enables automatic identification of the corresponding
patch for each vulnerability and supports direct interaction with
vulnerabilities after code changes, capabilities that existing large-scale
datasets do not provide. In our evaluation, ARVO successfully reproduces
81\% of vulnerabilities and achieves 89.4\% accuracy on the located
patches. We also discuss ARVO's influence on both upstream practices and
downstream security research.

\end{abstract}

\section{Introduction}

Vulnerabilities in software are both common and damaging: in 2024 alone, around 40,000 vulnerabilities were tracked by the National Vulnerability Database (NVD).
The prevalence of these flaws has spurred decades of research in automated vulnerability discovery and remediation, but historically, evaluation datasets to measure and understand these advancements have been elusive and ad hoc~\cite{klees2018evaluating}.

Databases which track vulnerabilities to alert users about security issues with software, such as the NVD, are primarily designed for system maintainers so they can verify affected versions and apply patches for known vulnerabilities in currently deployed software.
This mission does not naturally align with the needs of security research.
The NVD, and other datasets like it, typically lack critical metadata required for vulnerability research, including triggering inputs, reproducible environments, and ground-truth remediation data.
This hampers research: without a reproducible environment, it is difficult to measure the efficacy of dynamic analysis techniques, a lack of triggering vulnerability inputs hampers automated patch generation, and the verification of generated candidate patches is impossible without a ground truth reference.

Researchers have worked to produce a better dataset, but current approaches are insufficient.
Synthetic techniques, such as those represented by LAVA~\cite{LAVA} and challenge problems from the Cyber Grand Challenge~\cite{cybergrandchallenge_samples}, have questionable real-world overlap and implications.

Other techniques are more realistic, but require extensive manual effort leading to associated scaling problems.
For example, despite over 3,600 hours of human work to reproduce reported CVEs, Mu et al.~\cite{understand_reproducibility} only succeeded in reproducing a total of 368 vulnerabilities.
Other manually-crafted datasets show even greater scaling limitations: Magma contains just 7 projects with 118 unique vulnerabilities~\cite{Hazimeh:2020:Magma}.
This manual effort precludes continued generation of new data for these datasets, and when used as evaluation benchmarks, the datasets tend to become ``stale'' over time as researchers (perhaps unintentionally) tune their systems to eventually ``overfit'' on the data.

An emerging genre of dataset generation leverages large repositories of projects with historical vulnerability data to \emph{automatically} generate vulnerability datasets.
The most popular dataset is Google's OSS-Fuzz~\cite{serebryany_oss-fuzz_2017}, which provides findings from continuous fuzzing of over 1{,}000 open-source projects.
Unfortunately, repositories such as OSS-Fuzz were not designed to support this purpose: 63\% of the historical, vulnerable versions of software we attempted in our comparative evaluation fail to build due to errors caused by dependency drift, build system bitrot, and various exotic issues unaddressed by current techniques.
As a result, recent dataset generation techniques based on OSS-Fuzz, such as OSS-Fuzz-OSV~\cite{osv-source}, are able to reproduce only around 37\% of vulnerabilities.
Worse, even the vulnerability metadata in these datasets contains \emph{errors}: our research found 1,518 cases of incorrect vulnerability data (e.g., crashes or fixes) out of 10,440 records in OSS-Fuzz, an error rate of 14.5\%.
OSS-Fuzz-OSV inherits these issues downstream, resulting in a lower-quality dataset.

Full reproducibility, the ability to rebuild vulnerabilities from source code and reliably trigger the intended crash, has been absent from prior OSS (Open Source Software) security datasets.
We found that reproducibility is valuable for providing a verifiable feedback loop to downstream security research, such as patch locating and automated patch generation with large language models.
Building on this observation, we identify and solve the key challenges hampering building a full reproducible vulnerability datasets for security research.
Based on the solutions, we built \system\footnote{\url{https://github.com/n132/arvo}}, the ``Atlas of Reproducible Vulnerabilities.''
\system is both a framework designed to address the shortage of continually-updating, high-quality research vulnerability datasets and a comprehensive bug dataset in its own right.
\system aims to achieve a high level of reproducibility across a large number of real-world projects and vulnerabilities, providing a robust set of real-world vulnerabilities for a research vulnerability dataset.
We focus on memory-safety-related bugs in C/C++ projects that manifest as a crash on a proof-of-concept input.
We choose C/C++ for its widespread use and the significant impact of these bugs, though our methodology naturally extends to other languages.

\system has the following key capabilities: \begin{inparaenum}[(1)]
\item \emph{Recompilation}: \system can rebuild historical versions of
software afflicted by various vulnerabilities.
\item \emph{Precise fix identification}: \system automatically locates the precise developer patch that fixes the vulnerability.
\item \emph{Customization}: \system supports arbitrary modifications to reproduced vulnerabilities, enabling downstream applications such as the porting of vulnerabilities between software versions.
\item \emph{Triggering inputs}: Each vulnerability has a proof-of-concept ``triggering'' input that can be used to test for the presence of the vulnerability.
\item \emph{Accessibility}: We provide prebuilt container images for each vulnerability, allowing issues to be reproduced with a single command (e.g., \texttt{docker run -it arvo/example:42486945-vul arvo}).
\end{inparaenum}

\system works at a large scale, reproducing 6{,}138 vulnerabilities across 311 projects (81\% of attempted reproductions).
This represents a +44 percentage-point improvement over the state of the art ($\approx 2.19\times$) and, unlike the state of the art, effectively handles complex, multi-component software projects.
Additionally, we automatically evaluated developer-provided patches and confirmed that 89.4\% of these patches fix the corresponding vulnerability.
By cross-comparing \system and its upstream, \system identified and reported over 300 that were \emph{improperly} detected as fixed \emph{and remained unpatched to the present day}, and more that were incorrectly detected as fixed and remained unpatched for years.

Our results demonstrate that, in addition to reliably reproducing a high-quality vulnerability dataset for vulnerability research, \system's efficacy can uncover system issues in upstream components.
This has not gone unnoticed, and \system is already shaping both its upstream and its downstream.
\emph{Upstream}, Google, the maintainers of OSS-Fuzz, are currently finishing the process of integrating \system as OSS-Fuzz's vulnerability reproduction component.
\emph{Downstream}, \system served as a benchmark for \emph{multiple} teams in DARPA's AI Cyber Challenge (AIxCC)~\cite{aixcc}, and academic efforts such as CyberGym~\cite{wang2025cybergym} (an AI-agent cybersecurity benchmark) use \dataset as a data source.

\smallskip
\noindent
\textbf{\textit{Contributions.}} This paper makes the following
contributions:

\begin{enumerate}[leftmargin=*]
    \item
        We propose a new form of security dataset: introducing
        reproducibility to security datasets to build verifiable research
        vulnerability datasets at scale and with diversity.
    \item
    	We identify the key challenges in improving reproducibility for research vulnerability datasets and demonstrate our methods for addressing and mitigating these issues.
    \item
    	We design \system, a system that introduces reproducibility to security datasets, enables recompilation of historical vulnerabilities, and identifies more precise patch commits.
    \item
    	We present the \dataset, a reproducible, recompilable, and automatically updating dataset of over 6{,}000 real-world vulnerabilities in open-source C/C++ projects.
\end{enumerate}

Additionally, to support open research, we made \system itself---the framework, evaluation infrastructure, images, and metadata---\textbf{open-source}, so that other researchers can build on our work.
This includes 12{,}276 Docker images for reproducing each vulnerability (\texttt{vul} and \texttt{fix} versions for each of our 6{,}138 successful reproductions) and for re-compiling projects after any valid modifications of the source.

\section{Background and Related Work}

Security research depends heavily on security datasets. Two properties in
particular---quantity and diversity---have long been recognized as
essential as they enable more comprehensive evaluation. However,
despite significant effort and the existence of large-scale vulnerability
datasets, most vulnerabilities remain unusable, especially for
binary-focused research.

For example, the recent auto-patching work
PATCHAGENT~\cite{yu2025patchagent} was evaluated on 178 vulnerabilities
from 30 programs, while nearly 40{,}000 vulnerabilities were added to NVD
in a single year. Prior work~\cite{understand_reproducibility} highlighted
the gap between the vast number of discovered vulnerabilities and the
limited research-usable subset: reproducibility. By investing significant
manual effort, the authors made 368 vulnerabilities usable for security
research.

\subsection{OSS-Fuzz and OSV}
\label{sec:bk_OSS-Fuzz}

As the Internet's ``critical infrastructure''~\cite{cisa_oss}, open-source
software continues to gain prominence: as of Oct. 2024, GitHub hosted over
518 million~\cite{github_report:2024} public repositories. However, this
reliance on open-source software also amplifies the impact of security
vulnerabilities in widely used dependencies. One of the most impactful
efforts in this space is \emph{OSS-Fuzz}~\cite{serebryany_oss-fuzz_2017}, a
large-scale fuzzing project launched by Google. Since its launch in 2016,
OSS-Fuzz has continuously fuzzed more than 1{,}000 open-source projects,
finding and helping to fix over 10{,}000 vulnerabilities as of August 2025.
OSS-Fuzz monitors repositories, builds the software as new commits are
made, fuzzes with sanitizers (e.g., AddressSanitizer~\cite{asan}), reports
crashes, and periodically checks whether vulnerabilities have been fixed.

\emph{OSS-Fuzz-Vulns} is a subset of OSS-Fuzz that provides precise
impacted version ranges for its included vulnerabilities. The located
patches are provided to help software users determine whether their
deployed versions are affected by known issues. This dataset, integrated
into the Open Source Vulnerabilities project (OSV)~\cite{osv-source}, is
often referred to as OSS-Fuzz-OSV.

OSS-Fuzz-OSV is a semi-automated pipeline focused on providing precise
affected-version ranges: OSS-Fuzz-OSV avoids long-term reproduction
challenges by reproducing bugs immediately after they are reported in
OSS-Fuzz and archiving the located patches. This approach takes advantage
of recency: reproducing a crash right after discovery is much easier than
reproducing one from years earlier. It also biases OSS-Fuzz-OSV to
vulnerabilities with fewer dependencies, and the reproducibility is
therefore not maintained over time.

\begin{table*}[tb]
\centering
\caption{Prior existing datasets and comparison with ARVO. \emph{Type}:
how the dataset was constructed (Crafted = manually built; Generated =
automatically synthesized; Collected = aggregated from real projects).
\emph{Re-compilable}: each vulnerability can be rebuilt from source on
demand. \emph{Automated}: dataset construction does not require
per-vulnerability manual effort. \emph{PoC}: each vulnerability includes
a triggering input. \emph{Patch}: each vulnerability is linked to its
fix commit. \emph{Real-World}: vulnerabilities are derived from
real-world projects.}
\label{tab:dscompare}
\footnotesize
\begin{tabular}{l|c|r|r|c|c|c|c|c}
\toprule
\textbf{Dataset} & \textbf{Type} & \textbf{\# Vulns} & \textbf{\# Projects} & \textbf{Re-compilable} & \textbf{Automated} & \textbf{PoC} & \textbf{Patch} & \textbf{Real-World} \\
\midrule
CGC & Crafted & 276 & 249 & \cmark & \xmark & \cmark & \cmark & \xmark \\ 
ExtractFix & Crafted & 30 & 7 & \xmark & \xmark & \xmark &  \cmark & \cmark \\ 
Magma & Crafted & 118 & 7 & \cmark & \xmark & \cmark &  \cmark & \cmark \\ 
\midrule
FormAI & Generated & 112{,}000 & 1 & \xmark & \cmark & \xmark & \xmark & \xmark \\
D2A & Generated & 18{,}653 & 6  & \xmark & \cmark & \xmark & \cmark & \cmark \\
LAVA & Generated & 2{,}265 & 4  & \cmark & \cmark & \cmark & \xmark & \xmark \\
\midrule
Big-Vul & Collected & 3{,}754 & 348 & \xmark & \xmark & \xmark & \cmark & \cmark \\ 
PrimeVul & Collected & 6{,}968 & 755 & \xmark & \xmark & \xmark & \xmark & \cmark \\
DiverseVul & Collected & 7{,}514 & 295  & \xmark & \xmark & \xmark & \cmark & \cmark \\
CVEFixes & Collected & 5{,}495 & 1{,}754 & \xmark & \cmark & \xmark & \cmark & \cmark \\ 
OSS-Fuzz-OSV & Collected & 3{,}381 & 331 & \xmark$^{\dagger}$ & \hmark$^{\ddagger}$ & \cmark & \cmark & \cmark  \\ 
\midrule

\textbf{ARVO} & \textbf{Collected} & \textbf{6{,}138} & \textbf{311} & \cmark & \hmark$^{\S}$ & \cmark & \cmark & \cmark \\

\bottomrule
\end{tabular}
\vspace{0.3em}

{\raggedright $^{\dagger}$ The OSS-Fuzz-OSV data is collected via an
initial reproduction of each issue, but reproducibility is not guaranteed
over time. Reproduction is inherited from OSS-Fuzz rather than being solved
by OSS-Fuzz-OSV. Although OSS-Fuzz-OSV is a reproducible subset of
vulnerabilities of OSS-Fuzz, which implies fewer reproducibility issues, it
fails to reproduce complex (multi-component) or older vulnerabilities over
time. \par} {\raggedright $^{\ddagger}$ OSS-Fuzz-OSV is mostly automated; a
minority of reports are supplemented by manual input from project
maintainers. \par} {\raggedright $^{\S}$ ARVO's per-vulnerability pipeline
is fully automated; only a few hours of system-wide maintenance per season
are needed to redirect defunct external resources flagged by the pipeline.
\par}

\end{table*}
\vspace{\baselineskip}  

\subsection{Patch Locating}
\label{sec:patch_locating}
For known vulnerabilities, the corresponding source-code fixes, called
\emph{patches}, are vital to security research. Source code patches can be
used to detect their presence in
binaries~\cite{patch_detect_1,patch_detect_2,patch_detect_3} and to enable
hot-patch generation~\cite{hot_patch_1,hot_patch_2}. Revision control systems make patching possible by recording historical
changes. However, automatically identifying the relevant patch remains an
unsolved problem, even when a PoC is available. While
datasets such as CVE and NVD aim to alert system maintainers about
vulnerabilities and track affected software versions, they merely record
reported patches and do not verify or identify the actual fixing patch.

The available PoCs could, in principle, be used to test whether the
corresponding patch is present in a binary. In practice, however,
historical binaries are difficult to rebuild because existing
reproducibility solutions are limited, so PoCs can hardly be used to detect
the patch. Consequently, current methods typically rely on keyword matching
in commit messages and code. Automated methods such as
CVEfixes~\cite{bhandari_cvefixes_2021}, VCCFinder~\cite{VCCFinder}, and
PatchScout~\cite{patch_locating} are designed to map each CVE vulnerability
to its patch. Although they have shown good performance when evaluated on
CVE datasets, their accuracy is limited when it comes to locating patches
for non-CVE bugs. CVEs often receive considerable attention, making it
easier to gather detailed information about them, which aids in patch
identification. However, these methods struggle to generalize to the domain
of generic OSS bugs, which represent the majority of cases.

\subsection{Related Security Datasets}
\label{sec:priordatasets}

Existing datasets neglect reproducibility. Most focus on achieving either
quantity or diversity, while only a few datasets provide limited
reproducibility. To highlight these trade-offs, we review prior work across
three types of security datasets: crafted, generated, and collected. To
our knowledge, none achieves the combination of reproducibility, quantity,
and diversity that ARVO provides. Table~\ref{tab:dscompare} summarizes
\dataset with several representative security databases.
We divide these databases into three categories according to how they are
constructed and discuss their advantages and limitations.

\label{sec:data:subsec:compare}

\smallskip
\noindent
\textbf{Crafted Datasets} are manually built and are therefore often
limited in size. They are typically limited in both diversity and scale.
For example, the CGC dataset~\cite{cybergrandchallenge_samples} (276
vulnerabilities) was for evaluating automated vulnerability discovery
methods, while the ExtractFix~\cite{gao_beyond_2021} dataset (30
vulnerabilities) was for evaluating vulnerability patching. Some crafted
datasets also include high-quality real-world bugs.
Magma~\cite{Hazimeh:2020:Magma}, for instance, is a fuzzing benchmark
dataset that backports high-quality vulnerabilities into a recent version.
Nevertheless, their limited scale and manual, non-automated construction restrict their usefulness for binary security research.

\smallskip
\noindent
\textbf{Generated Datasets}, in contrast, are automatically created at a
large scale to synthesize or identify vulnerabilities. However, they often
lack diversity and complexity. For example, FormAI~\cite{FORMAI} produced
extensive datasets (112,000 vulnerabilities) but the average program code
length is only 79 lines, far simpler than real-world software. Although
FormAI covers over 60 CWE types, they come from only nine categories,
reflecting a
common limitation: generated datasets often lack diversity and focus on
similar vulnerabilities.

A second limitation is accuracy. D2A~\cite{D2A} achieves only 53\% label
accuracy, making it unreliable for research uses. To gain more accuracy,
LAVA~\cite{LAVA} takes a different approach by inserting bugs into existing
complex software and verifying reachability with PoCs. However, the
trade-off is poor diversity, as LAVA focuses solely on buffer overflows.

\smallskip
\noindent
\textbf{Collected Datasets}, unlike generated ones, are derived from
various real-world projects. This provides greater diversity and complexity
, but comes at the cost of scale, and reproducibility is particularly
challenging. As a result, these datasets often embody a trade-off between
quality (supported features) and quantity, and their limitations become
clearer when compared against one another.

Datasets such as Big-Vul~\cite{fan_cc_2020}, PrimeVul~\cite{PrimeVul},
DiverseVul~\cite{DiverseVul}, and CVEFixes~\cite{bhandari_cvefixes_2021}
offer a substantial number of vulnerabilities. Yet, none provide binaries
or re-compilation environments, making them unsuitable for Fuzzing or
auto-patching generation evaluation. They also face accuracy issues: in a
manual verification of labeling accuracy~\cite{DiverseVul}, DiverseVul's
labeling correctness is 60\%, while CVEFixes and Big-Vul have lower
accuracy rates of 51.7\% and 25\%, respectively.

In contrast, OSS-Fuzz-OSV prioritizes high-quality data over size. It
offers PoC--binary pairs, which support research using dynamic methods.
However, its binaries are \textbf{not recompilable}. Its official
reproducer struggles to recompile old or complex vulnerabilities, with an
overall success rate of 37\%. We return to these accuracy concerns in
Section~\ref{sec:comparison_osv}, where OSS-Fuzz is compared directly
against ARVO.

\smallskip
\noindent
\textbf{The ARVO dataset} that we present in this paper is, to our
knowledge, the \emph{first} to achieve large-scale reproducibility with
real-world software, combining quantity, diversity, and reproducibility.
We are not aware of any other public dataset achieving recompile-level
reproducibility at a comparable scale and across such a wide range of
projects. We further discuss the challenges
and benefits of reproducibility in Section~\ref{sec:reproducibility}, and
ARVO's solutions in Section~\ref{sec:design}.

\section{Reproducibility}
\label{sec:reproducibility}

As shown in Section~\ref{sec:priordatasets}, prior datasets emphasized
scale or diversity but neglected reproducibility, leaving researchers to
rely on only a small fraction of these datasets after substantial manual
effort to reproduce individual cases.

Achieving full reproducibility (rebuild plus retrigger) introduces several
challenges, which we highlight in the rest of this section.

\subsection{Reproduction for Vulnerabilities}

Reproducibility is the property that allows vulnerabilities to be reliably
reconstructed and studied, rather than remaining as text-based reports. It
depends on two complementary elements:

\smallskip
\noindent
\textbf{Reproducing Resources} are the metadata that
can help reproduce a vulnerability. These include vulnerability
descriptions, source code of the related components, the reproduction
environment, compilation methods or scripts, an example of a vulnerable
binary, the Proof of Concept (PoC) inputs that trigger the vulnerability,
and the corresponding patches. Not all of these resources are always
necessary, but each contributes to making reproduction more reliable.

\smallskip
\noindent
\textbf{Reproducing Pipeline} is the procedure that
uses available resources to consistently replay the expected behavior. An
effective pipeline should tolerate missing or partial resources while still
maintaining reliable success.

Prior work such as OSS-Fuzz-OSV connects \textbf{Reproducing Resources}
with the \textbf{Reproducing Pipeline} by reproducing a bug shortly after
it is reported and archiving the result, rather than maintaining the
ability to rebuild and trigger it on demand. As build environments and
dependencies drift, reproducibility degrades over time. What has been
missing is a methodology that sustains reproduction at scale, rebuilding
the vulnerable binary from source and triggering the intended crash
through the given PoC.

\subsection{Benefits}

Reproducibility enriches the information that can be extracted from
historical vulnerabilities. It improves data quality and enables downstream
researchers to reliably build and evaluate new security techniques.

\smallskip
\noindent
\textbf{Improving data quality.} Reproducibility makes it easier to filter
false positives and make the data verifiable, addressing a recurring
problem in large-scale collections. We quantify this in
Section~\ref{sec:app:subsec:zero-day}.

\smallskip
\noindent
\textbf{Enabling richer information extraction.} Reproducibility enables
extracting metadata that is otherwise hard to recover, such as the
vulnerability's fixing patch via commit bisection. We evaluate this in
Section~\ref{sec:comparison_osv}.

\smallskip
\noindent
\textbf{Supporting downstream research.} Current datasets still fall short
of the needs of downstream security research, such as program
repair~\cite{le_goues_genprog_2012, goues2019automated, yu2025patchagent}.
A reproducible \dataset covering a large number of real-world
vulnerabilities can be used to evaluate patch quality, and its
vulnerability-patch pairs can be integrated as training data for
machine-learning-based methods. Reproducibility also enables systematic
vulnerability backporting for fuzz testing. Because \dataset can rebuild
historical versions and revert fixing commits, it can automatically
backport known vulnerabilities into older software versions and verify
each vulnerability's reachability by its PoC. As we show in Section~\ref{sec:backporting},
this automation lets \dataset scale backporting across many projects,
complementing manually curated efforts such as Magma.

\subsection{Challenges}
\label{sec:challenges}
Constructing a large-scale, reproducible vulnerability dataset is far from
trivial. The primary difficulty lies in reviving historical software
versions, where compilation and environment setup frequently fail in subtle
ways. We summarize the key challenges below and describe ARVO's solutions
in Section~\ref{sec:design}, as well as a comprehensive ablation study in
Section~\ref{sec:ablation}.

\smallskip
\noindent
\textbf{Incompatible Dependencies.} Although historical
source code is usually preserved in revision control systems, recompilation
itself is a complex process. It typically requires many dependencies,
including not only specific libraries, but also certain versions of
compilers and system tools provided by the operating system. Open-source
software often depends on other components as well. For example, compiling
\texttt{ffmpeg} involves more than a dozen independent open-source
projects. If a reproducer ignores version control for these dependencies,
compilation will likely fail due to incompatible interfaces. Therefore, any
reproducible vulnerability dataset must capture and restore the exact
versions of its dependencies.

\smallskip
\noindent
\textbf{Missing Resources.} \label{sec:missing-resources} In
real-world scenarios, software compilation often requires fetching
resources from the Internet. These may include dependency source code
downloaded via \texttt{git clone} or special tools retrieved with \texttt{wget}
or \texttt{curl}. While this approach may work initially, over time the original
URLs often become invalid as projects migrate to new hosting platforms or
reorganize their repositories. For example, the PCRE library migrated from
an FTP-hosted SVN repository to a GitHub-hosted git repository, breaking
historical build scripts that relied on the old address. Since these
resources are essential for successful reproduction, any reproducible
dataset must capture or redirect them.

\smallskip
\noindent
\textbf{Fragile and Mercurial Build Processes.} The
build process is the most critical step in reproducing a vulnerability. To
keep the diversity and complexity from the upstream, we must handle
various build processes. Ideally, resource fetching and compilation should
be separated into distinct stages. In practice, however, historical build
scripts rarely follow this principle; build scripts commonly download
resources during compilation, which makes it harder to fix ``Incompatible
Dependencies'' and ``Missing Resources.'' Given the diversity of targets
and the complexity of real-world projects, this is particularly
challenging, as even minor modifications to the build process can easily
cause failures.

These challenges are not isolated; they often interact and amplify one
another, making reproduction far more complex than it initially seems. This
interdependence explains why prior datasets have largely ignored
reproducibility, despite its importance for research utility. Addressing
reproducibility in vulnerability datasets is therefore both novel and
non-trivial. In the next section, we introduce ARVO, which systematically
tackles these challenges through a scalable, automated design.

\section{ARVO}
\label{sec:design}

We designed \system with the goal of producing a reproducible and scalable vulnerability dataset that addresses the challenges mentioned in Section~\ref{sec:challenges}. In detail, we aim to achieve:

\medskip

\noindent
\textbf{Reproducibility.} ARVO should provide all the reproducing resources
(as Section~\ref{sec:challenges} discusses) along with a reliable pipeline
to recompile both vulnerable and fixed targets from source code.

\smallskip
\noindent
\textbf{Scalability.} The dataset should contain a large number of
vulnerabilities and automatically incorporate new vulnerabilities as they
are found, to allow the dataset to expand and grow easily over time.

\smallskip
\noindent
\textbf{Quality and Diversity.} Each vulnerability in the dataset should be
validated to ensure it is actually a bug with security impact. The
vulnerabilities should be distributed across a large number of different
projects to ensure that evaluations using the dataset are representative.

\smallskip
\noindent
\textbf{Ease of Use.} The dataset should be easy for researchers and
practitioners to use, without requiring extensive security expertise or
knowledge of how to build the projects.

\medskip
\noindent
In this section, we will describe the methods used in \system and how they mitigate the reproducibility challenges.

\subsection{Overview}

\system consists of two major components: (1) the reproducer and (2) the vulnerability patch locator. \system outputs the \dataset.

\system is a framework to generate a reproducible and interactive
research vulnerability dataset, designed to ingest source
metadata from ``bug''/project databases and augment this information with
relevant source code, build steps, and binaries. Because we hope to support
downstream uses such as analysis of security patches, evaluating
vulnerability discovery systems, and automated vulnerability repair, the
\dataset also needs to include environments for re-compiling the code of
each project so that modifications to the source code can be
straightforwardly tested. To enable easy access, \system provides an online
Dockerized dataset as well as infrastructure to build the dataset from
scratch.

The reproducer takes the provided metadata from the upstream bug
database(s), compiles the project at the specified (vulnerable) version,
and verifies that the provided triggering input causes a crash. It also
compiles the project at the fixed version listed in the upstream dataset
and checks that the crash no longer occurs. If either of these steps fails,
the vulnerability is marked unreproducible and excluded from the dataset.

However, as previously discussed, the upstream metadata often does not
specify the exact fixing commit, but only a range of candidates. \system's
vulnerability patch locator searches this commit range to find the earliest
commit that resolves the issue. ARVO's reproducibility enables the locator
to bisect the commit history and identify the exact changes that fix the
vulnerability.

\subsection{Upstream Dataset}
\label{sec:source-data}
To obtain a large number of vulnerabilities and allow the dataset to grow over time, \system is designed to draw project and bug metadata from upstream sources (currently, OSS-Fuzz). We rely on some assumptions about the upstream data source (discussed in Section~\ref{sec:reproducibility}):

\begin{enumerate}
  \item Version/Time Information: To reproduce and pinpoint its fix, we need version identifiers (e.g., git commit hashes) from the revision control system, which identifies the vulnerable and non-vulnerable versions of the project and its dependencies. When such identifiers are unavailable, ARVO falls back to using timestamps to approximate the correct versions, though this introduces some loss of precision.
  \item Build Environment: A virtualized, interactive
  environment that can compile and execute the target programs and their
  dependencies.
  \item Crash Information: At minimum, we need a triggering input and the command to execute the target program on that input. Additional information, such as sanitizer output, can also be used to validate that the crash we observe is the same one identified by the upstream source, though it is not strictly required.
\end{enumerate}

We chose OSS-Fuzz as our initial upstream source because of its diversity
and complexity. By introducing reproducibility into such a heterogeneous
dataset, we can more effectively demonstrate the robustness and generality
of ARVO's reproducibility solutions.

To identify security-relevant issues with the required metadata, we
searched the issue tracker according to the labels OSS-Fuzz automatically
applies to each issue: \texttt{Type=Bug-Security} (the crash is likely to
be security-relevant, based on the sanitizer report and call stack),
\texttt{label:Reproducible} (the crash occurs deterministically whenever
the triggering input is provided), and \texttt{status:Verified} (OSS-Fuzz
verified that the target no longer crashes).\footnote{We will see in
Section~\ref{sec:app:subsec:zero-day} that this label is not always
accurate; we found over 300 cases where the provided test case still
crashes the most recent version of the project.} Combining these query
elements, we obtain 8{,}921 issues in over 300 projects after filtering the
false positives (see Section ~\ref{sec:app:subsec:zero-day}), which serve
as the starting point for our dataset.

\subsection{Reproducer}
\label{sec:reproducer}

\begin{figure*}[t]
\centering
\includegraphics[width=0.9\textwidth]{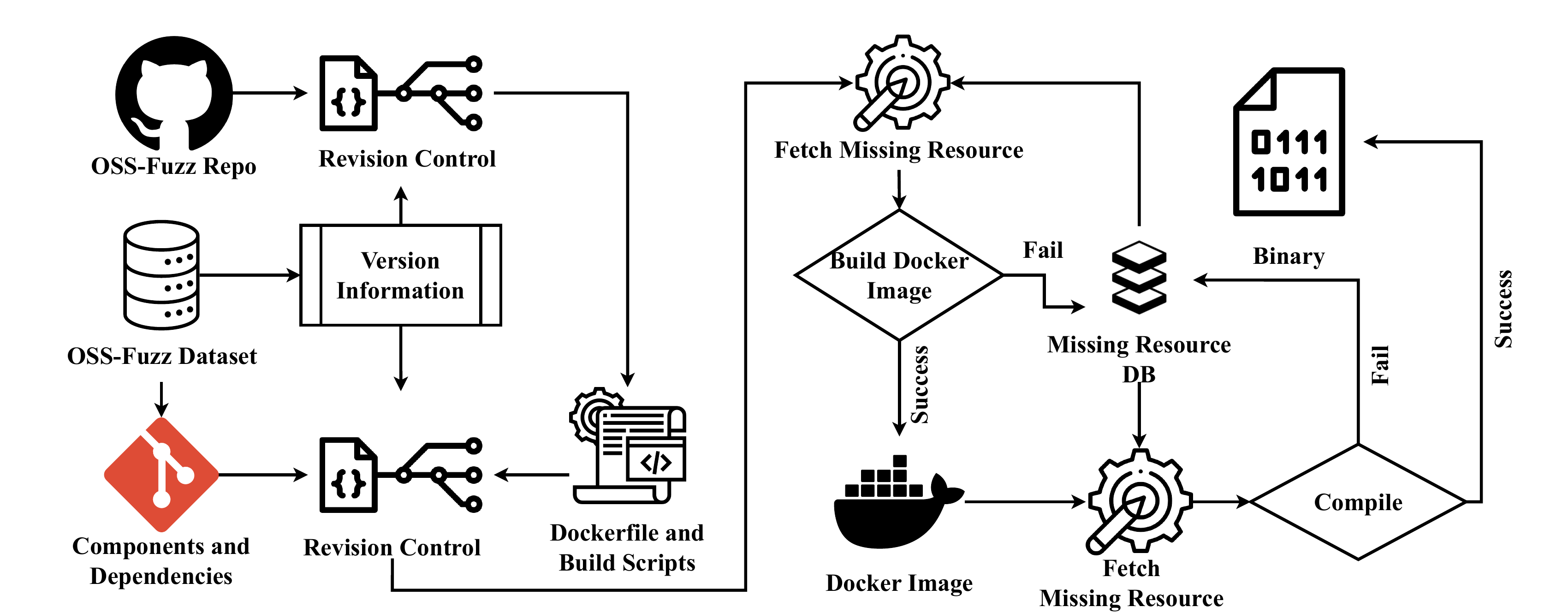}
\caption{ARVO Reproducer Structure. After reproduction, ARVO performs a verification step using the corresponding PoC: the vulnerable version must reproduce the intended crash, and the fixed version must run without crashing on the same input. Only verified cases are then packaged together with their build environment as a Docker image to support reproduction.}
\label{fig:reproducer}
\end{figure*}

The reproducer is the keystone of the patch locator: to reproduce an issue
and identify its precise fix, \system rebuilds the project and its
dependencies from source across different commits. This task is especially
challenging for older vulnerabilities, where dependencies, resources, and
toolchains may no longer be available.

By applying the techniques described in this section, the reproducer has
successfully reproduced 69\% of vulnerabilities (6{,}138 cases) with their
corresponding patches identified. The difference between this number and
the 81\% success rate observed in our comparison experiment (see
Section~\ref{sec:comparison}) is primarily due to time constraints. Some
OSS-Fuzz projects (e.g., LibreOffice) require several hours to compile, and
the repeated builds necessary for commit bisection make the process
extremely time-consuming. As a result, not all reproductions have been
completed; we expect the remaining gap to be closed in the coming months.

We identify three key strategies that \system uses to improve the
reproducibility of vulnerabilities: 1) minimally intrusive build
instrumentation; 2) revision control; and 3) fixing missing resources.
These strategies are implemented in the \system reproducer, as shown in
Figure~\ref{fig:reproducer}.

\smallskip
\noindent
\textbf{Build Instrumentation.}
Revision control during reproduction requires instrumentation of the build
process. OSS-Fuzz implements an intrusive revision control mechanism that
breaks the original compilation process. By contrast, ARVO applies
minimally intrusive revision control, preserving the natural build flow and
making reproduction more reliable.

\begin{figure}[t]
\centering
\includegraphics[width=\columnwidth]{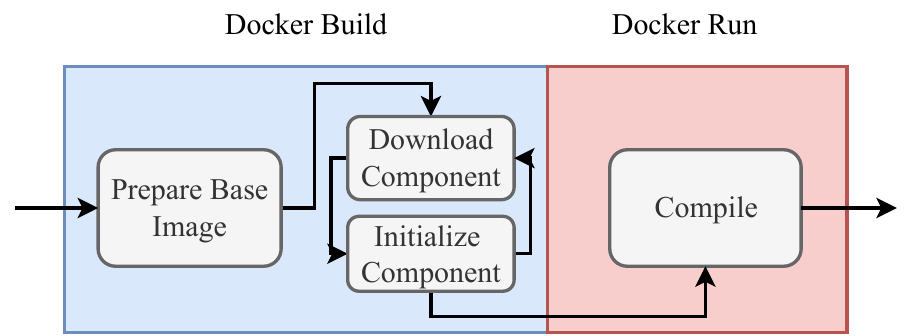}
\caption{Simplified Compilation Procedure.}
\label{fig:minimal_impact}
\end{figure}

OSS-Fuzz provides virtualized build environments by compiling projects
inside Docker. As shown in Figure~\ref{fig:minimal_impact}, the workflow
has two stages: Docker Build, which prepares resources, and Docker Run,
which performs compilation.

In OSS-Fuzz, revision control is enforced only after Docker Build
completes. The project source is swapped outside the container and mounted
back in before compilation. This approach ignores any resource
modifications or initialization steps that occur during the build stage.
Replaying these actions later is imprecise and risky, leading to broken
builds.

By contrast, ARVO sidesteps these issues by instrumenting the build process
with minimal change. ARVO separates actions into resource-download actions
and other actions. The revision control is only needed for resource
fetching actions, and ARVO only hooks these actions and keeps the
modification \emph{minimally intrusive}. The component initialization
actions during the Docker Build stage will not be changed.

In practice, ARVO hooks only resource-download commands instead of mounting
external sources. This minimally intrusive philosophy is also integral to
our approach to revision control, ensuring that modifications are both
effective and non-disruptive.

\smallskip
\noindent
\textbf{Revision Control.}
Correct dependency versions are essential for reproducing historical vulnerabilities, yet this requirement has long been overlooked. Existing efforts typically control only the main component's revision while ignoring its dependencies. However, successful reproduction depends on precise revision control of the entire build environment, including both the main project and all of its dependencies.

The build scripts used to compile the fuzz targets for each project are
provided by the project developers in two parts: a Dockerfile that
downloads dependencies and external resources, and a \texttt{build.sh}
script that actually compiles the fuzz targets. The official reproducer
provided by OSS-Fuzz rolls back the main project to the vulnerable commit,
but it does not attempt to reset dependencies to their corresponding
versions. This leads to compatibility issues and build failures since
dependencies have changed their APIs or build procedures. For instance,
\texttt{ImageMagick} relies on 15 separate components, each with frequently
changing APIs and usage patterns, and attempting to reproduce a
vulnerability in \texttt{ImageMagick} without adjusting the dependencies to
match the vulnerable version will likely result in a failed build. Also,
because the build script is usually attached to the main component and
separate from the dependencies, the old compiling commands in the build
script usually fail to compile the latest version of dependencies.

To demonstrate the impact of incorrect dependency versions, our ablation study in Section~\ref{sec:ablation} evaluates the effect of disabling component revision control.

\smallskip
\noindent
\textbf{Broken Resource Fixing.}
Similar to ``bit rot'' in software, reproducing old vulnerabilities often
encounters missing or inaccessible dependencies, especially for projects
from the 2017--2019 period. During this time, many projects migrated their
repositories from Subversion to Git, breaking build scripts that reference
the old repositories. ARVO mitigates this issue by maintaining a
detection-fixing loop that continuously identifies and fixes broken
resources.

ARVO divides missing resources into two categories: \emph{core resources},
which are necessary to compile the fuzz target, and \emph{non-core
resources}, which are not necessary for compilation but are required for
other parts of the build process. Core resources include software
dependencies, such as libraries, as well as tools used by the build process
that may be necessary to compile key components. Non-core resources include
documentation generation tools, seed/corpora used for fuzzing, and other
resources that are not directly related to the fuzz target.

To detect missing and broken resources, ARVO captures and logs the error
messages generated during the build process and looks for errors related to
failed URL downloads. It then classifies each missing resource as core or
non-core based on whether it is required to compile the fuzz target;
because ARVO performs reproduction rather than fuzzing, resources used
only by the fuzzing workflow can be safely skipped. For non-core resources,
such as corpora and seeds, ARVO modifies
the related commands to prevent build crashes, with specific
modifications depending on the command type to ensure minimal loss. For
core resources, limited manual work (a few hours per season) is required to
locate the missing resource and replace it with a working URL. \footnote{An
experimental version of \system uses LLMs to reduce manual effort, primarily
by resolving compilation issues and filtering upstream false positives. The
results reported in this paper do not depend on any LLM involvement.}
These
resource fixups are stored as reusable rules that can be applied across
multiple vulnerabilities and projects.

Although the manual effort might appear daunting, in practice, most missing
resources are shared across many vulnerabilities and projects. Fixing a
relatively small set of resources, therefore, resolves a large number of
cases. For example, our dataset includes only 53 unique missing resources
identified over the past eight years; keeping ARVO's fixes updated would
thus require updating roughly seven records per year. This translates to
just a few hours of work annually, yet it enables substantial gains in
reproducibility, allowing us to successfully reproduce thousands of
additional vulnerabilities

\subsection{Patch Locator}

Automatically locating vulnerability patches, as discussed in
Section~\ref{sec:patch_locating}, is nontrivial even when a PoC is
available because of the absence of a reliable reproducer. With its
reproducer, \system can automatically locate the precise patches that
resolve each vulnerability. This section introduces \system's patch
locator, which applies commit bisection to identify fixing commits. We
first describe how bisection is used to locate patches in the main
component, and then extend the discussion to the more challenging case
where dependencies must also be tracked during bisection.

\begin{figure}[tb]
  \centering
  \includegraphics[width=0.9\columnwidth]{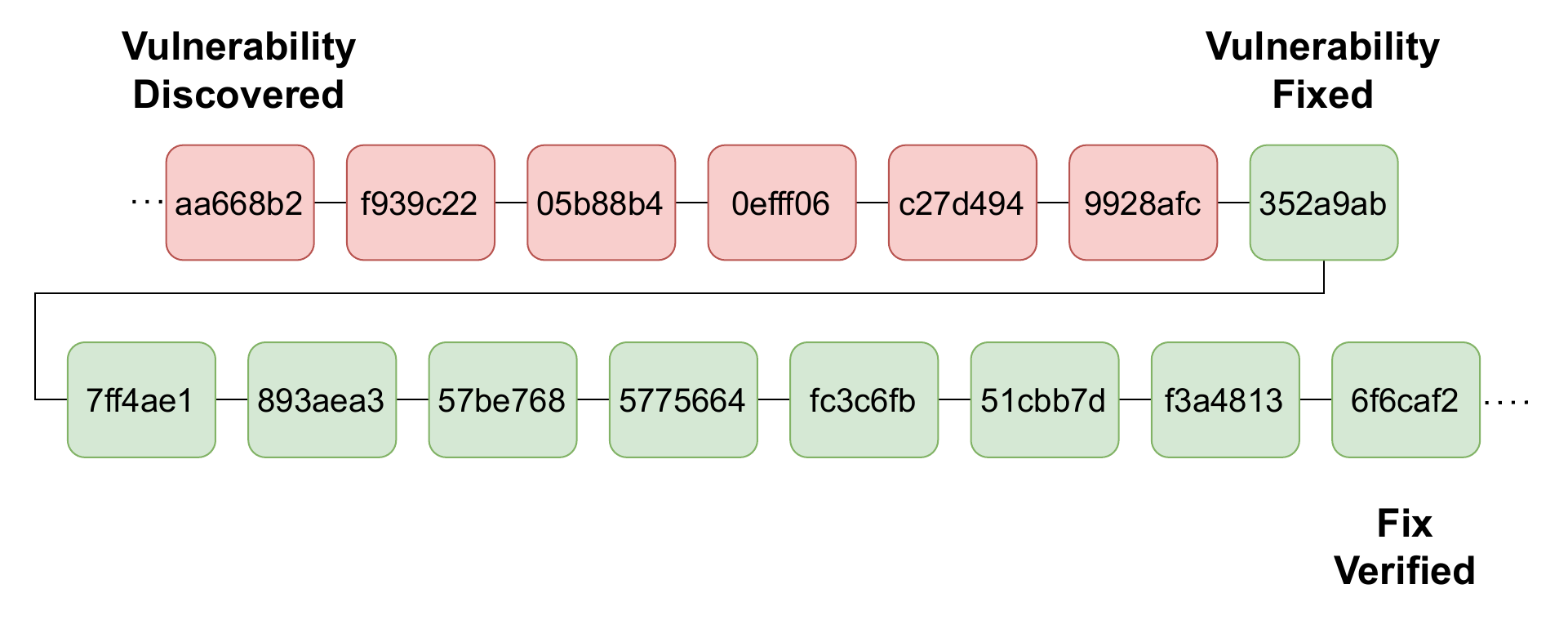}
  \caption{Vulnerability lifecycle for OSS-Fuzz issue~\#42508698 on \texttt{ImageMagick}.}
  \label{fig:issue42508698}
\end{figure}

Benefiting from the combination of continuous fuzzing and data collection,
OSS-Fuzz stands out because it automatically provides, for each
discovered bug, a range of commits that includes the patch. The reported
result is a range rather than a single commit because OSS-Fuzz checks for
vulnerability fixes periodically rather than commit by commit. For most projects, this is typically a daily build. During
their daily builds, OSS-Fuzz checks whether the latest version still
crashes on the recorded PoC. If it does not, the issue is marked as fixed.
For highly active projects, however, the delay between the maintainers' fix
and OSS-Fuzz's verification can create a range of candidate commits for the
actual patch. There are also cases that include larger ranges, where
OSS-Fuzz did not verify whether the patch PoC was applied for a longer time.

Figure~\ref{fig:issue42508698} illustrates the process through a concrete
example: the vulnerability is first identified at
\texttt{6f6caf}; the maintainers commit a fix the next day at
\texttt{aa668b}; and OSS-Fuzz verifies the fix at \texttt{aa668b}. However,
\texttt{aa668b} is not the actual fix but merely a minor update to the
\texttt{ChangeLog} file. To identify the actual fix, we must search
over the 14 commits and 83 files that were changed between the initial
report and the verification.

\system's reproducer makes it possible to use commit bisect to locate
the vulnerability patch. However, once dependencies are taken into account,
the task becomes more complex than simply bisecting the main component's
commits.

During bisection, every chosen commit of the main component must be
compiled and tested with the triggering input. To keep these builds
working, \system also needs to identify the correct dependency versions for
each commit. If the dependencies are mismatched, the project may fail to build at a
chosen midpoint. \system handles such cases by stepping to a neighboring
commit so that bisection can proceed. In the extreme case where a high
fraction of intermediate commits fail to build, bisection can no longer
make progress and \system falls back to a \emph{linear search}: it walks
candidate commits one by one, which is significantly more expensive and
may yield a range of commits rather than a single fix commit. Unlike in reproduction, the fix locator cannot rely on
dependency versions recorded upstream, since those versions may no longer
apply across the entire commit range.

To address this, \system uses commit timestamps. For each main component
commit, it selects the most recent compatible dependency commits available
at that time, ensuring that the project remains buildable throughout the
bisection process.

\begin{figure}[tb]
\centering
\includegraphics[width=0.405\textwidth]{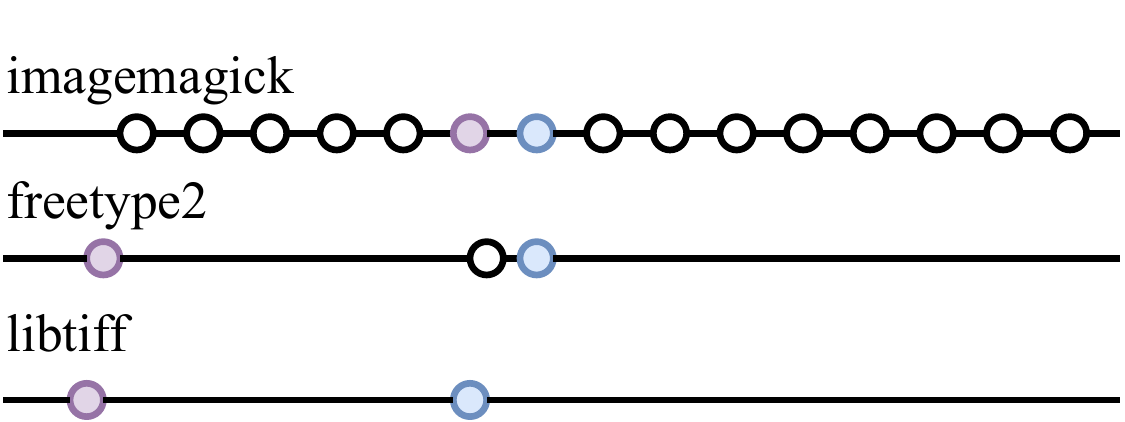}
\caption{Revision control for the locator. To compile the blue/purple
revision of \texttt{ImageMagick}, corresponding dependency revisions are selected
using commit timestamps. Only 2 of 14 dependencies are shown for clarity.}
\label{fig:vc_locator}
\end{figure}

Figure~\ref{fig:vc_locator} illustrates this approach with the \texttt{ImageMagick}
case. For each revision of the main component, we use the commit timestamp
to select the appropriate dependency versions, which greatly improved build
reliability. With this precise version control in place, ARVO can bisect
across thousands of commits to accurately pinpoint the fixing commit.

\subsection{Beyond OSS-Fuzz}
\label{sec:beyond-ossfuzz}

Although this paper focuses on OSS-Fuzz, the methodology described above
is not specific to it: any upstream that records vulnerable source
revisions, build instructions, and triggering inputs can be plugged in.
As a proof of generality, we ported the pipeline to a different upstream,
syzbot~\cite{syzbot}, which reports kernel bugs found by syzkaller. Each syzbot report
supplies a syz-language reproducer (and often a C reproducer) along with
the affected kernel commit, which maps directly to ARVO's required inputs.

We sampled 144 syzbot reports at random; 44 referenced commits in
orphan linux-next or subsystem trees that were no longer reachable from
mainline and thus fell outside the methodology's scope. On the remaining
100 reachable reports, the prototype reproduced 78 (78\%); 20 built and
ran but did not crash within the timeout (typically race conditions),
and 2 failed to build. The reproduced
crashes span \texttt{KASAN} reports, \texttt{BUG}, \texttt{WARNING}, and general protection faults.

The Linux kernel is in some respects a friendlier target than OSS-Fuzz:
it is a single, well-maintained codebase with a uniform build system, so
most of ARVO's original challenges (heterogeneous build scripts, missing
third-party resources) do not apply. On the other hand, kernel bugs
include a substantial proportion of race conditions that resist PoC-based
reproduction, which is the dominant cause of the 20 ``built but did not
crash'' cases above. The OSS-Fuzz evaluation in
Section~\ref{sec:case-study} therefore remains the more demanding and
informative one; the kernel result only confirms that ARVO's solutions
transfer to a non-OSS-Fuzz upstream.

\subsection{Dataset Access}

A goal of our dataset is that it should be easy to use, even for
researchers who do not have a security background; we hope that this will
allow researchers in other fields (e.g., machine learning) to use it as an
evaluation target. Based on \system, we have uploaded Docker images for
each vulnerability to Docker Hub, enabling each issue to be reproduced and
recompiled with a single command:\\
{\footnotesize \texttt{docker run <repo name>:<localId>-<vul|fix> arvo
[compile]}.}

To support more advanced uses of the dataset (e.g., rebuilding the project
with other instrumentation), we open-source \system so
researchers can rebuild the \dataset from scratch with their desired
changes.

\section{Dataset}
\label{sec:dataset}
This section presents the details of the \dataset constructed using the methods described in Section~\ref{sec:design}.

\subsection{Dataset Characteristics}
\begin{figure}
\centering
\includegraphics[width=1\columnwidth]{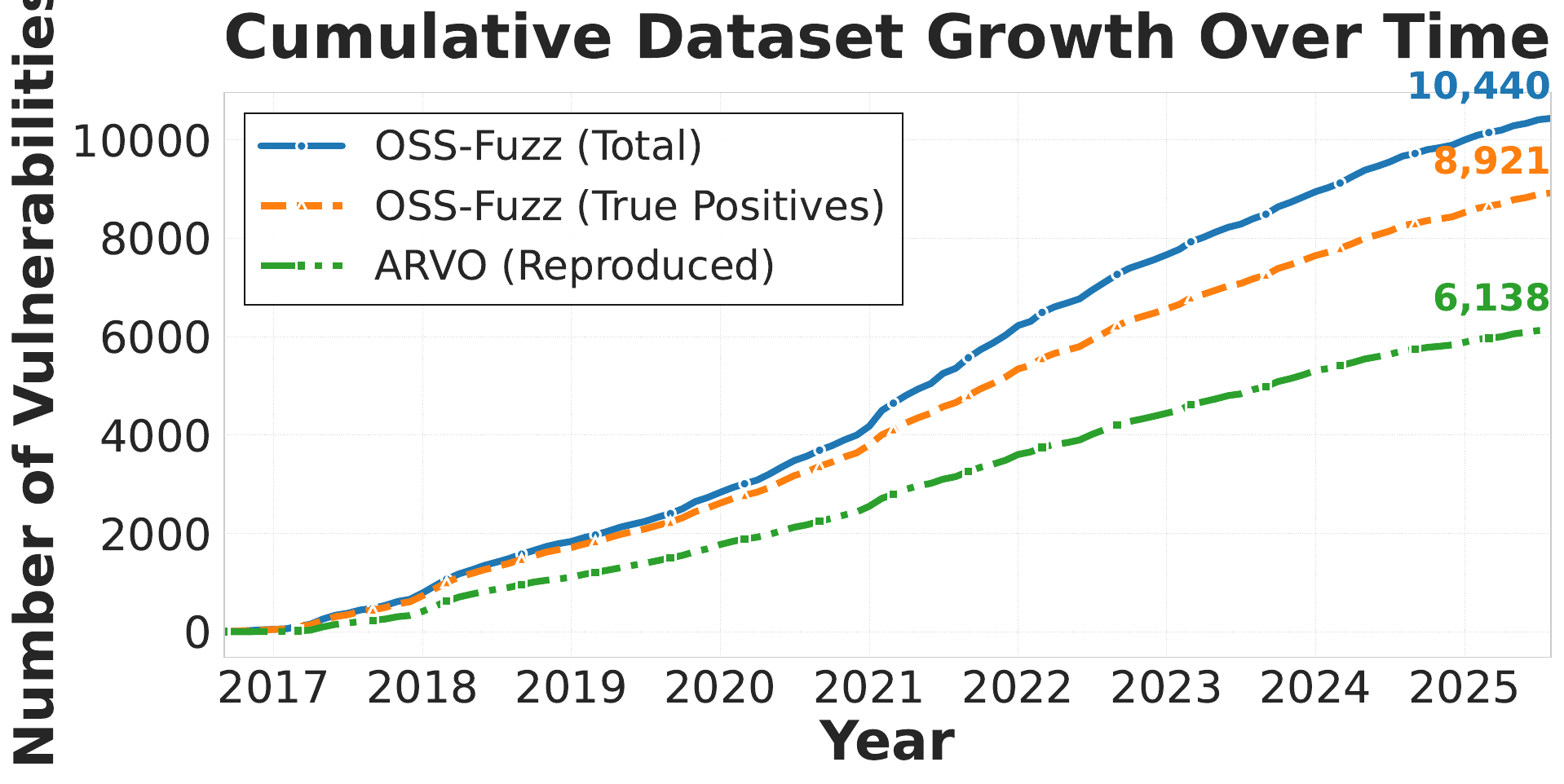}
\caption{Database growth over time.}
\label{fig:growth}
\end{figure}

\smallskip
\noindent
\textbf{Dataset Size and Growth.} At the time of
writing, \system has successfully reproduced 6{,}138 vulnerabilities across
311 projects, out of the 8{,}921 vulnerabilities initially obtained from
OSS-Fuzz. For each reproduced case, we provide interactive environments
supporting recompilation and instrumentation for deeper analysis. In
addition, 221 of these vulnerabilities are also linked to their
corresponding CVE identifiers.

We measured our reproduction success over time and found that \system has
maintained a roughly constant success rate (Figure~\ref{fig:growth}). Our
design emphasizes a general approach that works across diverse open-source
projects. As Figure~\ref{fig:growth} shows, the size of the \dataset grows
steadily in tandem with OSS-Fuzz, which continues to expand its project
coverage and discover new bugs. By leveraging its upstream's ongoing
efforts, \system is well positioned to scale into the future, ensuring that
the \dataset continues to grow and stay up to date.

\smallskip
\noindent
\textbf{Project and Language Distribution.} To
demonstrate the diversity of \dataset, we computed the distribution of
vulnerabilities among 311 projects. This distribution is relatively even;
the top 10 projects collectively account for only 35.71\% of all vulnerabilities in the dataset.
This indicates the comprehensive diversity of the \dataset: Rather than
concentrating around a small set of projects, it spans a wide range of
software.

\smallskip
\noindent
\textbf{Duplicates.} In \dataset, ``duplicates'' are
kept. During upstream fuzzing, a vulnerability could crash on different
harnesses/functions by different fuzzing engines (e.g., libfuzzer and AFL).
These ``duplicates'' are crashes triggered by different PoCs but fixed on the
same commit. Based on \system located patches, we can connect these crashes to the same cause. ARVO keeps them in the dataset but marks them as different crashes fixed by the same patch. Preserving and labelling duplicates in this way is preferred for downstream applications, such as auto-patch generation, since more PoCs could be used to verify the correctness of generated patches.

\smallskip
\noindent
\textbf{Patch Statistics.} Of the 6{,}138
vulnerabilities in the \dataset, we first filtered out duplicates (e.g.,
when a single patch fixed multiple vulnerabilities---1{,}550 cases).

The large size of this dataset allows us to collect some interesting
statistics on the nature of vulnerability patches. Prior research has found
that security-related fixes are typically small and
self-contained~\cite{li_large-scale_2017}. Our data also supports this
finding: In \dataset, 90\% of the patches modified 4 files or fewer; 2{,}895 patches (63.22\%) affect just
a single file. Looking at the number of lines added and removed by each
patch, we found a median of 6 lines added and 2 lines removed; the mean of
both is significantly larger (228.7 added and 115.1 removed) due to a small
number of outliers. 90\% of the patches in our dataset have fewer than 62
lines added and fewer than 32 lines removed.

We also found that 53.68\% (2{,}458) of the patches directly modify the
source file and function present in the sanitizer-reported call stack. Of
these, 37.46\% of the patches have modified the function on the
$0^{\text{th}}$ index in the crashing call-stack, 32.88\% of the patches
modified functions at the $1^{\text{st}}$ index, and 25.49\% of the patches
modified functions on the $2^{\text{nd}}$ index.

\subsection{Data Accuracy Evaluation}
\label{sec:comparison_osv}
To evaluate the accuracy of \dataset, we compare it directly against
OSS-Fuzz-OSV. By manually analyzing random samples from both ARVO and
OSS-Fuzz-OSV, we obtain an overview of patch correctness and the relative
reliability of each dataset.

\smallskip
\noindent
\textbf{Overview.} \dataset has 6{,}138 successfully
identified patches while OSS-Fuzz-OSV has documented 3{,}381
vulnerabilities with patches. To make the comparison, we focus on the
overlapping cases between them, a total of 2{,}219 vulnerabilities. The
overlapping set of vulnerabilities has 0.83 dependencies on average, while
the average \dataset vulnerability has 4.05 dependencies. Based on the
located patch commits, we divided them into three groups: agree (66.61\%),
disagree (20.91\%), and partially agree (12.48\%).

We manually evaluated 100 random cases from each group. In agree and
partially agree cases, 97\% and 93\% were confirmed as true positives,
respectively. Most partially agreed cases involve a merging commit and an
effective commit in another branch. We considered a case a true positive
when the overlapping commit patched the vulnerability. There were a total
of 7 false positives in these two groups, and all of them were due to
changes in the harness or compilation settings rather than actual
vulnerability fixes.

In the disagree group, where ARVO and OSS-Fuzz-OSV reported different
patches, the accuracy drops. Specifically, 63\% of ARVO's results were true
positives (versus 32\% for OSS-Fuzz-OSV).

Based on the group sizes and per-group accuracies, the overall accuracy is
computed as a weighted average:
\[
\text{Overall accuracy}
= \frac{\sum_{g \in \{\text{agree},\,\text{partial},\,\text{disagree}\}} N_g \, p_g}
       {\sum_{g \in \{\text{agree},\,\text{partial},\,\text{disagree}\}} N_g},
\]
where \(N_g\) is the size of group \(g\) and \(p_g\) is its accuracy.
Applying this formula yields 89.4\% for \dataset, outperforming
OSS-Fuzz-OSV's 82.9\% on patch data accuracy.

Both ARVO and OSS-Fuzz-OSV show high correctness in the agree and partially
agree groups. However, in the disagree group, ARVO achieves nearly double
the accuracy rate compared to OSS-Fuzz-OSV. In the 100 randomly sampled
cases, ARVO exclusively provided valid patches for 46 vulnerabilities,
whereas OSS-Fuzz-OSV identified only 15 such cases. In addition, there were
17 cases where both systems located valid patches, and 22 cases where
neither system provided a valid patch.

\smallskip
\noindent
\textbf{Analysis.} Although OSS-Fuzz-OSV benefits from
maintainer input when identifying fix commits, \system achieves higher
accuracy. The key advantage lies in reproducibility.

Unlike ARVO, the OSS-Fuzz-OSV dataset includes not only patches confirmed
by binary search but also maintainer-reported fixes, which can introduce
false positives. For example, in issues~\#42529818 and~\#42486491, OSS-Fuzz-OSV
linked the vulnerabilities to plausible but ultimately unrelated patches.
After extensive investigation, we found that these patches did not address
the target vulnerabilities, whereas \system correctly identified the
functional fixes. Reproducibility also helps improve patch quality. In
issue~\#42541392, ARVO's recursive binary search pinpointed the actual fix in
a submodule, whereas OSS-Fuzz-OSV listed only a submodule update commit.

However, this bisection-based patch locating method also has limitations. A
PoC failing to trigger a crash does not necessarily mean that the
vulnerability is fixed. Any change that makes the bug unreachable, such as
modifications in the fuzzing harness or compilation settings, may suppress
the crash without addressing the root cause. Automated upstream datasets
like OSS-Fuzz may accept such changes as valid fixes, which can also
mislead ARVO. Our manual evaluation shows that ARVO’s results are highly
reliable (89.4\%), with the remaining \(\sim10\%\) false positives,
primarily due to patches that suppressed the crash without actually fixing
the underlying vulnerability. Overcoming this challenge will require
combining dynamic methods (bisection) with complementary techniques such as
semantic understanding. By ensuring reproducibility, ARVO provides the
foundation for such future advances.

\section{Case Studies}
\label{sec:case-study}
This section presents three case studies that illustrate both the
effectiveness of ARVO's reproduction solution and its broader value for
security research. Together, they demonstrate how reproducibility benefits
ARVO itself, its downstream applications, and even its upstream datasets.

The first case study examines ARVO directly, using an ablation study to
measure the contribution of its core features to reproducibility. The
second demonstrates a downstream use: leveraging ARVO to reintroduce known
bugs into software and construct reliable fuzzing benchmarks. The third
shifts upstream, analyzing cases where ARVO failed to reproduce
vulnerabilities and showing that many of these failures stem from OSS-Fuzz
false positives rather than limitations of ARVO. Collectively, these
studies highlight not only ARVO's strengths but also the broader impact of
reproducible datasets across software security research.

\subsection{Reproducibility Comparison and Ablation Study}
\label{sec:comparison}

To demonstrate ARVO's improvement in reproducibility and to highlight the
significance of its core contributions, we compare ARVO with OSS-Fuzz's
official reproducer and conduct an ablation study on ARVO's key solutions.

\smallskip
\noindent
\textbf{Reproducibility Comparison.} Due to recent
upstream changes (OSS-Fuzz migrated its issue tracker), ARVO had to rebuild
its entire dataset. Reproduction and patch localization are time-consuming
because they require multiple rounds of compilation. Consequently, the
6{,}138 completed cases do not reflect ARVO's full capacity, since
it has not finished reproducing all upstream issues. To provide a fair
and controlled comparison, we randomly sampled 100 vulnerabilities from
OSS-Fuzz. Each issue was reproduced using both the official OSS-Fuzz
reproducer and the ARVO reproducer.

\begin{table}[tb]
\caption{Successful Reproduction Counts for 100 Random Issues.}
\label{tab:enable_disable}
\centering
\footnotesize
\begin{tabular}{c c c c}
\toprule
\textbf{\# Components} & \textbf{\# Total} & \textbf{\# OSS-Fuzz} & \textbf{\# ARVO} \\
\midrule
1     & 47  & 27 & 41  \\
2--4  & 29  & 10 & 19  \\
5--10 & 10  &  0 &  9  \\
$>$10 & 14  &  0 & 12 \\ \midrule
\textbf{Total} & \textbf{100} & \textbf{37} & \textbf{81} \\
\bottomrule
\end{tabular}
\end{table}

As shown in Table~\ref{tab:enable_disable}, ARVO achieves a success rate of
81\%, substantially outperforming OSS-Fuzz's 37\%. OSS-Fuzz succeeds mainly
in projects with a single component and struggles in multi-component
projects due to its lack of dependency version control. This indicates that
the OSS-Fuzz-OSV dataset is biased toward simpler projects, overlooking
more complex ones. In contrast, ARVO consistently improves reproducibility
across both simple and complex projects, covering 67 distinct projects in
total. The 81\% success rate demonstrates that ARVO's solutions effectively
mitigate the key challenges in vulnerability reproduction.

\label{sec:ablation}

\smallskip
\noindent
\textbf{Ablation Study.} To evaluate the contribution
of ARVO's key features to reproducibility, we conducted an ablation study
by selectively disabling (1) Revision Control for non-main components, (2)
Resource Fixing, and (3) Base Environment Control (providing the
corresponding base image environment for compilation). The results are
shown in Table~\ref{tab:ablation}.

\begin{table}[tb]
\caption{Reproduction Counts with ARVO Features Disabled.}
\label{tab:ablation}
\centering
\footnotesize
\begin{tabular}{l c c c}
\toprule
\textbf{Disabled Feature} & \textbf{\# Reproduced} & \textbf{\# Lost} & \textbf{Success Rate} \\ \midrule
RF              & 54 &  27 & 66.7\% \\
BE              & 50 &  31 & 61.7\% \\
RC              & 46 &  35 & 56.8\% \\
RF and BE       & 46 &  35 & 56.8\% \\
RC and RF       & 42 &  39 & 51.9\% \\
RC and BE       & 37 &  44 & 45.7\% \\
RC, RF, and BE  & 34 &  47 & 42.0\% \\
\bottomrule

\end{tabular}

\vspace{0.3em}
RF = Resource Fixing, BE = Base Environment, RC = Revision Control.
\end{table}

In the ablation study, disabling each feature reduced ARVO's success rate
to varying degrees. Disabling only Resource Fixing has a smaller effect,
while both Full Revision Control and Matching Base Environment caused
larger drops. When features were disabled in combination, reproduction
success decreased sharply, showing that all three components contribute in
complementary ways.

Overall, ARVO improves reproducibility substantially compared to OSS-Fuzz
(81\% vs. 37\%). The ablation results confirm that ARVO's features are not
only individually useful but also mutually reinforcing. Together, they
enable ARVO to scale to more complex projects and provide a strong
foundation for downstream vulnerability research.

\subsection{Vulnerability Backporting}
\label{sec:backporting}

\begin{table}[b]
\centering
\caption{Vulnerability Backporting Result (Top 5)}
\label{tab:IntroBugs}
\footnotesize
\begin{threeparttable}
\begin{tabular}{l r r}
\toprule
\textbf{Project Name} & \textbf{\# Verified Vulns\tnote{a}} & \textbf{\# Affected Files} \\
\midrule
ghostscript & 45 & 27 \\
assimp & 43 & 26 \\
mupdf & 30 & 26 \\
selinux & 14 & 11 \\
opensc & 13 & 10 \\
\midrule
\textbf{Total} & \textbf{145} & \textbf{100} \\
\bottomrule
\end{tabular}
\begin{tablenotes}
\scriptsize
\item[a] The triggered vulnerabilities after deduplication
\end{tablenotes}
\end{threeparttable}
\end{table}

ARVO and Magma share a goal of building fuzzing benchmarks from real vulnerabilities, but pursue it through different strategies.
This case study demonstrates how ARVO enables the creation of such benchmarks at scale, complementary to (rather than directly comparable with) Magma's approach.
Recent benchmark efforts such as Magma highlighted the importance of realistic bugs for evaluating fuzzing performance.
Magma forward-ports past vulnerabilities into a recent version of the software, which is valuable for evaluating mitigations and fuzzers against modern codebases, but requires manual patch adaptation; Magma originally introduced 118 bugs into 7 projects, later expanding to 9.

ARVO automates Magma's approach.
Rather than relying on manual patch adaptation, ARVO searches among the commits related to each vulnerability (e.g., the commits where the bug was first detected and later fixed) for a target commit.
By default, ARVO selects the commit on which most bugs can be triggered, maximizing benchmark size.
If the user instead requires a specific recent commit (matching Magma's evaluation scenario), ARVO supports that target commit as well, at the cost of fewer triggerable bugs.

Once a target commit is chosen, ARVO compiles the program with varying combinations of inserted vulnerabilities and verifies each inserted bug by its PoC, keeping only the bugs whose PoC still triggers the intended crash.
The resulting benchmark therefore contains only reachable bugs, together with their PoCs and corresponding patches.

To evaluate this idea, we applied it to 34 projects, excluding complex projects whose compilation takes hours and the ones with fewer than 40 historical vulnerabilities.
For each project, we search for the code version suitable for porting most known vulnerabilities.
As shown in Table~\ref{tab:IntroBugs}, the project inserted most vulnerabilities in our evaluation contains 45 verified vulnerabilities, each confirmed by its corresponding PoC.
The crash reports from these PoCs cover 9 distinct crash types, including \texttt{global-buffer-overflow}, \texttt{heap-buffer-overflow}, and \texttt{heap-use-after-free}.
The automatically backported vulnerabilities are also not concentrated in a small set of files but are spread across the codebase: in the top 5 projects alone, 145 vulnerabilities land in 100 different files.
This is achieved without any manual patch adaptation across versions (avoiding patch-application issues), at a scale beyond what manual approaches such as Magma can practically reach across many projects.

While this case study demonstrates the potential of automated vulnerability insertion, it also highlights interesting challenges that invite further exploration.
Reverting a fix to reintroduce a bug onto a different commit faces the same core challenge as patch backporting: when the codebase has changed substantially between the bug's original location and the target commit, the patch (or its reverse) no longer applies cleanly.
Even when a patch applies successfully, peripheral changes to surrounding code can leave the patched code referencing dependencies (functions, types, or fields) that no longer exist, causing compilation to fail.
In our experiments on the \texttt{ndpi} project, most attempts failed for this reason.
The similar challenge is discussed in patch-porting research~\cite{shariffdeen_patch_transplant_2020,shariffdeen_fixmorph_2021}.
We did not integrate these techniques in our evaluation, and applying them to ARVO is left for future work.
Another challenge lies in triggerability: not all reintroduced vulnerabilities can be triggered by their associated PoCs.
Shallow and easy-to-trigger bugs (e.g., \texttt{Use-of-uninitialized-value}) can prevent PoCs from reaching deeper vulnerabilities, masking the intended crash.
As a concrete example, we ran an exhaustive backporting attempt on the \texttt{ghostpdl} project, which contained 177 historical vulnerabilities at the time of the experiment. \system successfully backported 45 of them; among the 132 failures, 63 were deep vulnerabilities blocked by shallower ones reaching the crash first, 27 were patches that could not be cleanly reverse-applied, and 42 were cases in which the original PoC no longer triggered the vulnerability.

Despite these limitations, the case study shows that ARVO can insert historical vulnerabilities into diverse projects more efficiently than previous approaches, demonstrating both its scalability and effectiveness.
\subsection{Correct False Positives on the Upstream}
\label{sec:app:subsec:zero-day}

Accurate data is the foundation of reliable research. Conclusions drawn
from noisy or incorrect vulnerability reports risk being misleading. While
analyzing cases that ARVO failed to reproduce, we found that nearly half
were not due to ARVO's limitations, but rather to incorrect upstream data.
In these instances, ARVO could build the software, yet the provided PoCs
failed to demonstrate the expected behavior: either the pre-patch commit
did not crash, or the post-patch commit still did. This revealed two
issues: numerous OSS-Fuzz reports correspond to false positives, and many
patches are incorrectly labelled as fixes.

We identified \textbf{1{,}519 false positives} by combining ARVO's
reproduction results with a confirmation step against the original
binaries archived by OSS-Fuzz: when even running the upstream-archived
binary on its recorded PoC failed to reproduce the expected crash, the
vulnerability was treated as a false positive rather than as an ARVO
reproduction failure. These cases fall into three categories: (1) crashes
that cannot be triggered consistently (e.g., race conditions or OSS-Fuzz
internal bugs), polluting the dataset; (2) crashes fixed outside the commit
range identified by OSS-Fuzz, leading to mislabeled patches; and (3)
crashes still reproducible on the latest upstream commit, posing a serious
risk since attackers could exploit these unfixed vulnerabilities.

Handling category~(1) is not complex: such crashes can be filtered out to
avoid unreliable data. For category~(2), ARVO leverages its unique reproducing
ability to locate the true fix by bisecting the commits between the latest
version and the commit where the crash was first observed. As an
illustrative example,
issue~\#42486945~\cite{ossfuzz42486945}
on OSS-Fuzz was claimed fixed. However, OSS-Fuzz pointed to a
commit~\cite{libheif_readme_commit}
that merely modified a README file as the ``fix'' of the heap buffer
overflow, which is clearly incorrect. With ARVO, we located the actual
patch~\cite{libheif_actual_patch}
(Appendix~\ref{appendix:patch42486945}), which was applied nearly two years
later. The commit message indicates that it fixes a wrong size passed to
\texttt{memcpy}, and according to the crash call stack reproduced by ARVO,
the modified function is indeed responsible for allocating the overflowed
heap chunk. Furthermore, ARVO's de-duplication revealed that
issue~\#42502614 shares the same fix. Although different fuzz engines
(\texttt{honggfuzz} and \texttt{afl}) were used, the crash reports were
nearly identical.
Thus, issue~\#42486945 remained an unfixed vulnerability for around
\textbf{two years}, with OSS-Fuzz incorrectly labelling it as resolved
while pointing attackers directly to the PoC.

Finally, for category~(3), we confirmed and reported more than \textbf{300
cases} to the upstream, where the very latest commit remained vulnerable.
By feeding these corrections back, ARVO strengthens the reliability of
upstream datasets and avoids leaking information about unfixed
vulnerabilities.

\section{Discussion}

\system presents a novel type of vulnerability dataset by introducing \emph{reproducibility} into scalable upstreams, turning static records into interactive artifacts that can be rebuilt from source.
The resulting \dataset offers comprehensive vulnerability data to support security research.
However, it primarily serves as a mitigation to recover lost reproducibility.

Upstream infrastructures typically archive the crashing binary but omit the \emph{entire build environment} and other reproduction details.
This missing context makes it increasingly difficult to reproduce vulnerabilities over time.
A more sustainable long-term approach is for upstream sources to capture and preserve reproducibility information directly.
If the build environment and related context were archived, datasets like \system would not need to reconstruct them later, and the broader open-source security community would benefit from easier, more reliable reuse of vulnerability data.
Even if \dataset itself becomes unnecessary once upstreams adopt reproducibility, the central contribution of this paper remains: introducing reproducibility into security datasets.

\subsection{Limitations}
Even though \system demonstrates significant improvement over prior
datasets, the methodology does face certain limitations.

\smallskip
\noindent
\textbf{Source-Dependent.} ARVO relies on upstream
datasets and their metadata. While \system can detect some issues, broken
metadata from the upstream dataset may lead to false positives.

\smallskip
\noindent
\textbf{Vulnerability Scope.} ARVO is biased toward
vulnerabilities that manifest as a crash on a PoC input. Bugs that are
hard to reproduce, such as race conditions whose triggering window is
narrow, are excluded; so are vulnerability classes without an observable
crash signal (e.g., logic flaws). Extending the methodology to these
classes is left for future work.

\smallskip
\noindent
\textbf{PoC-based Reproduction.} While reproducing
vulnerabilities, we did not enforce strict matches on the crash type and
address, which may affect the accuracy of the \dataset (i.e., it is
possible that the crash we reproduce differs from the original
vulnerability, despite sharing the same triggering input).

\smallskip
\noindent
\textbf{Patch Quality.} \system's reliance on
bisection for identifying vulnerability fixes has limitations. Due to the
possibility of multiple related commits, this approach might not always
accurately pinpoint the exact fix, particularly when fixes involve a series
of modifications. Also, while a commit may lead us to a ``correct'' fix, an
individual commit might encompass extensive modifications, complicating the
identification of the precise change responsible for the fix.

\smallskip
\noindent
\textbf{Time-Consuming.} Even with Docker images
provided, re-compiling downstream applications remains costly. Although
running the commands is straightforward, rebuilding the entire \dataset
from scratch is resource-intensive: our most recent rebuild, including
patch locating, took roughly 4 weeks on a 192-core server with 256\,GB of
RAM. Patch locating further amplifies this cost, as failed commits can
force linear search instead of bisection.

\smallskip
\noindent
\textbf{Duplicated Cases.} The \dataset contains cases
where upstream reported multiple vulnerabilities that share the same
underlying root cause. While these duplicates may appear redundant, they
are valuable for patch verification, as their PoCs trigger crashes through
different execution paths. We therefore preserve and label them rather than
discarding them, to better test patch robustness. However,
patch-commit-based labeling remains imprecise, since a single patch may
simultaneously fix multiple vulnerabilities.

\subsection{Future Work}

The reproducer of ARVO has officially merged into OSS-Fuzz.
In the near future, we will merge the patch locator into OSS-Fuzz and correct the corrupted metadata on OSS-Fuzz to provide cleaner vulnerability information.
In parallel, we will continue to maintain \system as an open-source project to fill the gap until upstream sources enforce reproducibility natively, at which point a separate \system would no longer be needed.
To broaden its scope, we also plan to incorporate additional upstream sources, starting with the Linux kernel vulnerabilities, where more vulnerability information can be retrieved from the mailing history and commit messages.

Looking ahead, we will focus on making the reproduction process more reliable, improving the accuracy of \dataset, and making \system easier to use as a dependable foundation for future security research until the day full reproducibility is introduced into upstream.

We also plan to extend ARVO's bug-insertion feature with prior patch-porting techniques~\cite{shariffdeen_fixmorph_2021,shariffdeen_patch_transplant_2020} to increase the number of bugs that can be successfully reintroduced onto a single target commit, and to explore LLM-assisted variants for cases where source-code changes between the bug's origin and the target commit defeat purely syntactic approaches.

\section{Conclusion}

In this paper, we present ARVO, an Atlas of Reproducible Vulnerabilities in
Open-source software, establishing reproducibility as a core property for
security datasets. We identified key challenges in vulnerability
reproduction and proposed practical solutions, transforming
document-centric reports into a reproducible database with interactive
environments. Our open-source dataset and framework include more than
6{,}000 vulnerabilities across 311 projects, represented in over 12{,}000
interactive build images hosted on Docker Hub. Accuracy evaluation shows
that 89.4\% of located patches are correct, underscoring the reliability of
our approach. By combining reproducibility, scale, and diversity, ARVO
offers the most comprehensive dataset of its kind to date, with a framework
designed to automatically incorporate new vulnerabilities and projects in
the future.

\section*{Open Science}
\label{sec:open_science}
All artifacts necessary to reproduce the results in this paper are
publicly available:

\begin{itemize}
    \item \textbf{Source code:} \url{https://github.com/n132/arvo}
    \item \textbf{Dataset:} \url{https://github.com/n132/ARVO-Meta}
    \item \textbf{Evaluation data:} \url{https://github.com/sefcom/ARVO}
\end{itemize}
\section*{Ethics Considerations}
\label{sec:ethics}
We have found more than 300 potentially unfixed bugs and 2381 potential
false positives, which we reported to OSS-Fuzz, and further responsible
disclosure for the existing bugs is being processed. It is worth noting that
throughout our experiments we only used publicly available data from
the OSS-Fuzz issue tracker (fixes, PoCs, etc.), i.e. we have not found any new
vulnerabilities.

\section*{Acknowledgments}

We would like to thank the anonymous reviewers for their insightful feedback.

This material is based upon work supported by NSF 2247954, NSF 2146568, NSF 1663651, NSF 2442984, and NSF 2232915, by the Defense Advanced Research Projects Agency (DARPA) under Young Faculty Award D22AP00145-00, and by generous support from the US Department of Defense.
This work is also sponsored by, and related to, Department of Navy award N00014-23-1-2563 issued by the Office of Naval Research, and supported in part by the Australian Government through the Australian Research Council's Discovery Projects funding scheme (project DP250101396).
Hammond Pearce is supported in part by a Google Research Scholar award.

Any opinions, findings, and conclusions or recommendations expressed in this material are those of the author(s) and do not necessarily reflect the views of the National Science Foundation, the Office of Naval Research, the Department of Defense, DARPA, or any other sponsor. The content of the information does not necessarily reflect the position or the policy of the Governments, and no official endorsement should be inferred.

\bibliographystyle{unsrt}
\bibliography{ref/benhamram}

\appendix
\lstset{
    basicstyle=\small\ttfamily, 
    frame=single,               
    breaklines=true             
}

\section{Located Patch 42486945}
\label{appendix:patch42486945}

\begin{lstlisting}[caption={Patch for Open Issue 42486945},label={appendix:42486945_patch}]
commit 11ffeffadd980f9f96019fe180fc1e81827e3790
Author: Dirk Farin <dirk.farin@gmail.com>
Date:   Mon Apr 4 20:43:45 2022 +0200

    fix wrong memcpy size

diff --git a/libheif/heif_colorconversion.cc b/libheif/heif_colorconversion.cc
index 2b05068..5a07ebb 100644
--- a/libheif/heif_colorconversion.cc
+++ b/libheif/heif_colorconversion.cc
@@ -526,7 +526,8 @@ Op_YCbCr_to_RGB<Pixel>::convert_colorspace(const std::shared_ptr<const HeifPixel
     }
 
     if (has_alpha) {
-      memcpy(&out_a[y * out_a_stride], &in_a[y * in_a_stride], width * 2);
+      int copyWidth = (hdr ? width * 2 : width);
+      memcpy(&out_a[y * out_a_stride], &in_a[y * in_a_stride], copyWidth);
     }
   }
\end{lstlisting}

\end{document}